\documentclass[preprint,prl,aps,amsmath,amssymb,color,superscriptaddress]{revtex4}
\usepackage{stmaryrd}
\usepackage{amsmath}
\usepackage{amssymb}
\usepackage{graphicx}
\usepackage{dcolumn}
\usepackage{bm}
\usepackage{tabularx}
\usepackage{color}
\begin{document}
\begin{titlepage}
\title{Spin-Triplet Excitonic Insulator: The Case of A Semi-Hydrogenated Graphene}
\author{Zeyu Jiang}
\affiliation{State Key Laboratory of Low-Dimensional Quantum Physics and Collaborative Innovation Center of Quantum Matter, Department of Physics, Tsinghua University, Beijing 100084, China}
\author{Wenkai Lou}
\affiliation{SKLSM, Institute of Semiconductors, Chinese Academy of Sciences, P.O. Box 912, Beijing 100083, China}
\author{Yu Liu}
\affiliation{Laboratory of Computational Physics,Institute of Applied Physics and Computational Mathematics, Beijing 100088, China}
\author{Yuanchang Li}
\email{yuancli@bit.edu.cn}
\affiliation{Key Lab of advanced optoelectronic quantum architecture and measurement (MOE), and Advanced Research Institute of Multidisciplinary Science, Beijing Institute of Technology, Beijing 100081, China}
\author{Haifeng Song}
\affiliation{Laboratory of Computational Physics,Institute of Applied Physics and Computational Mathematics, Beijing 100088, China}
\affiliation{CAEP Software Center for High Performance Numerical Simulation, Beijing 100088, China}
\author{Kai Chang}
\affiliation{SKLSM, Institute of Semiconductors, Chinese Academy of Sciences, P.O. Box 912, Beijing 100083, China}
\affiliation{Beijing Academy of Quantum Information Sciences, Beijing 100193, China}
\author{Wenhui Duan}
\affiliation{State Key Laboratory of Low-Dimensional Quantum Physics and Collaborative Innovation Center of Quantum Matter, Department of Physics, Tsinghua University, Beijing 100084, China}
\affiliation{Institute for Advanced Study, Tsinghua University, Beijing 100084, China}
\author{Shengbai Zhang}
\affiliation{Department of Physics, Applied Physics and Astronomy, Rensselaer Polytechnic Institute, Troy, NY, 12180, USA}
\date{\today}

\begin{abstract}
While various excitonic insulators have been studied in the literature, due to the perceived too-small spin splitting, spin-triplet excitonic insulator is rare. In two-dimensional systems such as a semi-hydrogenated graphene (known as graphone), however, it is possible, as revealed by first-principles calculations coupled with Bethe-Salpeter equation. The critical temperature, given by an effective Hamiltonian, is 11.5 K. While detecting excitonic insulators is still a daunting challenge, the condensation of triplet excitons will result in spin superfluidity, which can be directly measured by a transport experiment. Nonlocal dielectric screening also leads to an unexpected phenomenon, namely, an indirect-to-direct transition crossover between single-particle band and exciton dispersion in the semi-hydrogenated graphene, which offers yet another test by experiment.
\end{abstract}

\maketitle
\draft
\vspace{2mm}
\end{titlepage}
Macroscopic quantum phenomenon like superconductivity is relatively rare but rather attractive both from a fundamental point of view and from its potential technological applications. As an analogue to the BCS superconductors, excitonic insulator was first proposed more than 50 years ago \cite{Kohn}, which has a reconstructed many-body ground state as a result of the exciton binding energy ($E_b$) exceeding the single-particle band gap ($E_g$). It corresponds to a state characterized by a spontaneous Bose condensation of excitons\cite{Halperin,KohnRMP}. So far, however, compelling experimental evidence for such an intriguing state is still lacking although much progress has been made in recent years\cite{Kogar,Du,Lu}. This is in part because the requirement of $E_b > E_g$ is often difficult to meet in solid state and in part because the excitons are charge-neutral. Their uniform flow may not cause an electrical current, which hinders the detection of the exciton condensate by transport measurement as in the case of superconductivity.

Generally, as $E_g$ decreases, the system screening increases so $E_b$ decreases synchronously\cite{usPRL}. To realize $E_b > E_g$, ones need to break this synergy between $E_g$ and $E_b$ by eliminating band-edge transitions\cite{usEI,usHEI}. In systems with weak spin-orbit coupling, electric-dipole transitions obey the spin selection rule which can also serve as a solution to this problem. More importantly, spin-forbidden here will force electron-hole excitations with spin flip, giving rise to spin-triplet ($S$=1) excitons. Unlike the spin-singlets ($S$=0), the flow of triplet excitons carry spins. A spin superfluidity has been predicted when they condense into a single quantum state\cite{SunQF}, just like a $^3$He\cite{Leggett}. In turn, the superfluidity straightforwardly signifies an exciton condensation and can be directly measured by a spin transport experiment.

Owing to exchange interactions, triplet excitons are omnipresent in solids\cite{Khaliullin,KunePRB,Kaneko}. However, to probe the spin transport, a macroscopic spin-polarization might be required, suggesting that one should look for excitonic instability in magnetic systems. An electronic structure with the top valence band and bottom conduction band of opposite spins is suitable here, as it will lead to and only lead to spin-triplet excitons. In condensed matter, indeed, there are ample examples where the band edge states exhibit such a desired feature. Below, we will concentrate on two-dimensional (2D) systems, for their significantly enhanced $E_b$ and noticeably decreased exciton radius as a result of the reduced screening. Both are beneficial to the observation of exciton condensation\cite{Rubio}. A good example is the semi-hydrogenated graphene, known as graphone\cite{ZhouJ} as shown in Fig. 1(a).

In this paper, first-principles electronic structure calculations reveal excitonic instability for spin-triplet excitation in graphone. In addition, the calculated exciton dispersion unveils that the zero momentum exciton is energetically most stable, despite that graphone has an indirect single-particle band gap. One can explain such a counterintuitive result from the nonlocal screening properties of the host material, which map out the response of the electron-hole pairs over the entire Brillouin zone. While the finding here may not be restricted to 2D, the unique electronic properties of graphone may certainly magnify such an effect. Being a direct-transition excitonic insulator also protects the graphone from a major obstacle in experimentally identifying the exciton condensation, namely, the intervention of Jahn-Teller distortion in indirect-gap materials\cite{Kogar}. Using an effective Hamiltonian model, we further find a many-body gap of 24.1 meV for the graphone in the excitonic-insulator phase, which leads to a critical temperature around 11.5 K.

Our electronic structure calculations employed the density functional theory within the Perdew-Burke-Ernzerhof (PBE) exchange-correlation functional\cite{PBE}, Heyd-Scuseria-Ernzerhof (HSE06) hybrid functional\cite{HSE06} and many body $GW$ approaches\cite{Hybertsen} in the VASP code\cite{vasp}. The PBE was used to find the geometric and magnetic configurations of the ground state. A plane-wave cut-off of 70 Ry was used. An $18 \times 18 \times 1$ \emph{k}-point grid was used to sample the Brillouin zone. To determine the magnetic ground state, both lattice constants and atomic positions were fully relaxed until residual forces were less than 1 meV/\AA. Quasiparticle $GW$ calculations\cite{Hybertsen}, which include single-shot $G_0W_0$ and partially self-consistent $GW_0$, were carried out on top of PBE (denoted as $GW_{0}$@PBE and $G_{0}W_{0}$@PBE, respectively). For the exchange and correlation parts of the self-energy, we used the energy cut-offs of 50 and 15 Ry, respectively. A total of 192 bands were used to ensure an $E_g$ convergence to within 0.01 eV. Excitonic properties were obtained by solving the Bethe-Salpeter Equation (BSE) \cite{Rohlfing} on top of the $GW$ results with only the top valence and bottom conduction bands included. Formation energy and wave functions for the \textbf{q}=0 excitons were calculated by the Yambo code\cite{yambo}, while the full exciton dispersion was obtained by using the Exciting code\cite{exciting}. When performing the BSE calculations for PBE band structures, we have corrected $E_g$ to the $GW$ values by a scissor operator. This scheme has been successfully applied to study various excitonic effects\cite{usPRL,LiYang} and has predicted an excitonic instability in low-dimensional materials\cite{usEI,usHEI,Varsano}.

Table I lists the calculated minimum (indirect) $E_g$, as well as the direct gap at $\Gamma$, of the graphone. $E_g$ increases from 0.53 eV (PBE), 1.91 eV (HSE06), 3.01 eV ($G_0W_0$@PBE), to 3.49 eV ($GW_0$@PBE). For diamond and silicon, the $G_0W_0$@PBE method yields a reasonably $E_g$ close to experiments\cite{ChenW}. This method also reproduces the experimentally observed optical gap of graphane\cite{Yang,usPRL}, which is a double-side hydrogenated graphene. Besides, the $G_0W_0$@PBE method predicts a gap between the highest occupied molecular orbital (HOMO) and the lowest unoccupied molecular orbital (LUMO) for the C$_8$H$_8$ molecule, which is 0.36 eV higher than experiment\cite{Govoni}.
In this study, we will use the $GW_0$@PBE method, as in the literature this method is often considered to be accurate, but judging from the above results, it may actually slightly overestimate the gap of graphone.

\begin{figure}[tbp]
\includegraphics[width=0.75\columnwidth]{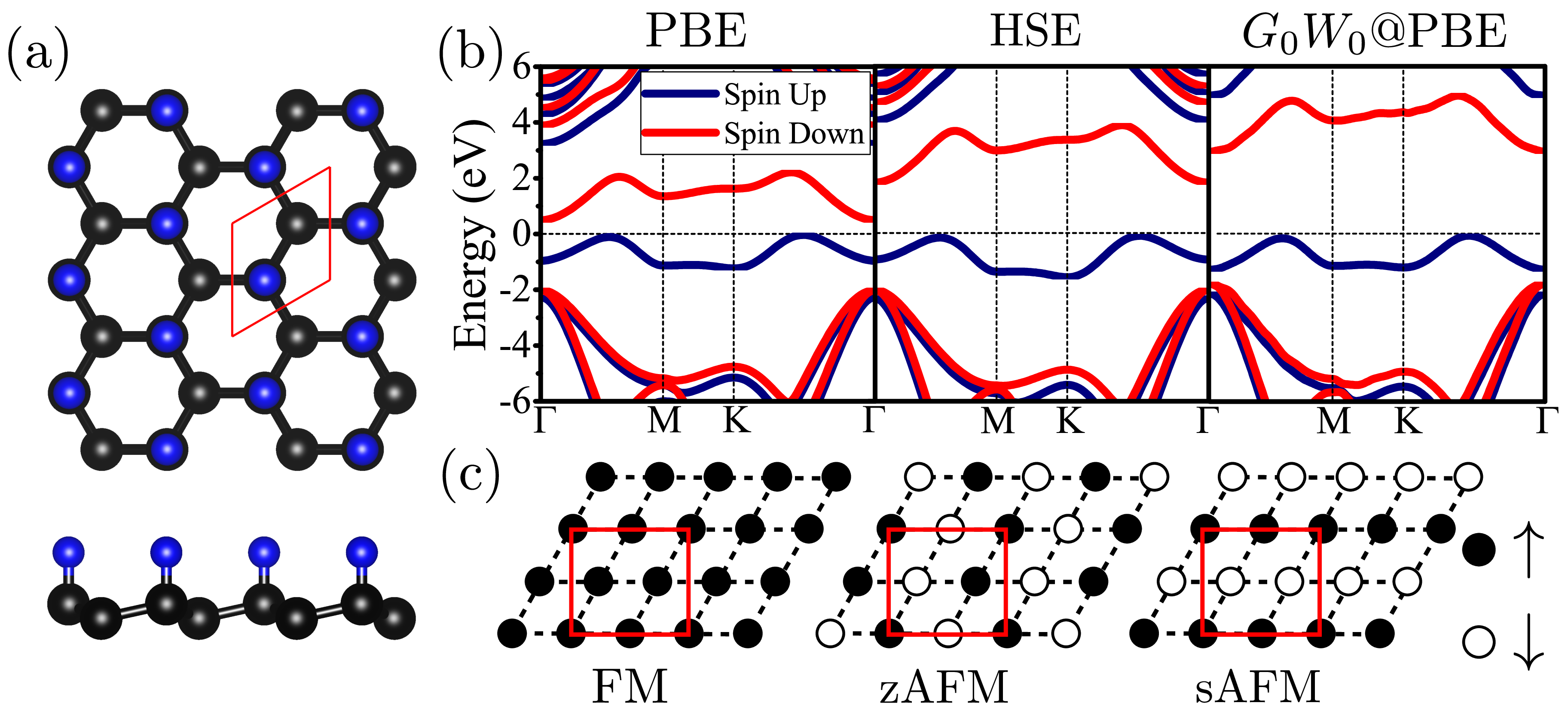}
\caption{\label{fig:fig1} (Color online) (a) Top and side views of the graphone with carbon (hydrogen) atoms colored in black (blue). Red rhombus denotes the unit cell. (b) Band structures within different calculating methods. The valence band maximums are set as energy zero (Fermi energy). (c) Magnetic configurations considered: ferromagnetic (FM, left), zigzag (zAFM, middle), and stripe antiferromagnetic (sAFM, right). Each circle represents one unsaturated C atom, empty and solid to distinguish spin moments of opposite direction. Red rectangles are the supercells used for the energy evaluation.}
\end{figure}

Figure 1(a) shows the atomic structure of graphone where an alternative hydrogenation of the $p_z$ orbitals breaks the $\pi$-bonding network. Geometrically, this leads to a buckling of the graphene basal plane. Electronically, it causes the unsaturated $p_z$ orbitals to split into two bands: one is below zero (occupied) and one is above zero (empty) as can be seen in Fig. 1(b). Each unsaturated $p_z$ orbital is expected to produce a local magnetic moment of 1 $\mu_B$. We have calculated different magnetic structures and found that the ferromagnetic state is the ground state and is lower in energy by 31 and 39 meV (per H), respectively, relative to the zigzag and stripe antiferromagnetic states in Fig. 1(c). These results are in line with previous study\cite{ZhouJ}.

\begin{table}
\centering
\caption{Direct and indirect single-particle band gaps and formation energy ($E_f$) of the lowest-energy excitation (= $X_1$ exciton) given by different calculation methods in unit of eV.}
\renewcommand\arraystretch{1.0}
\begin{ruledtabular}
\begin{tabular}{lcccccccccccccccccccccccccc}
        & PBE & HSE & $G_0W_0$@PBE &$GW_0$@PBE \\
\hline
   Direct gap & 1.46 & 2.76 & 4.22 & 4.64 \\
   Indirect gap & 0.53 & 1.91 & 3.01 & 3.49 \\
   $E_f$ (\textbf{q}=0) &   -3.35 & -2.00 & -0.55 & -0.13 \\
\end{tabular}
\end{ruledtabular}
\end{table}

Figure 1(b) shows single-particle band structures for graphone by three different methods. On the appearance, they look all similar with the conduction band minimum at $\Gamma$ and the valence band maximum between $\Gamma$ and $K$. Indeed, the topmost valence band and the bottommost conduction band possess opposite spins, which is of particular importance: on the one hand, the spin-selection rule prevents electric dipole transition between them, which has been shown to be critically important for decoupling $E_g$ with $E_b$ and subsequently the creation of an excitonic instability\cite{usPRL,usEI,usHEI}. On the other hand, the exciton involved in these two bands must be spin-triplet, whose coherent flow can thus be experimentally measured to verify the presence of exciton condensation.

\begin{figure}[tbp]
\includegraphics[width=0.75\columnwidth]{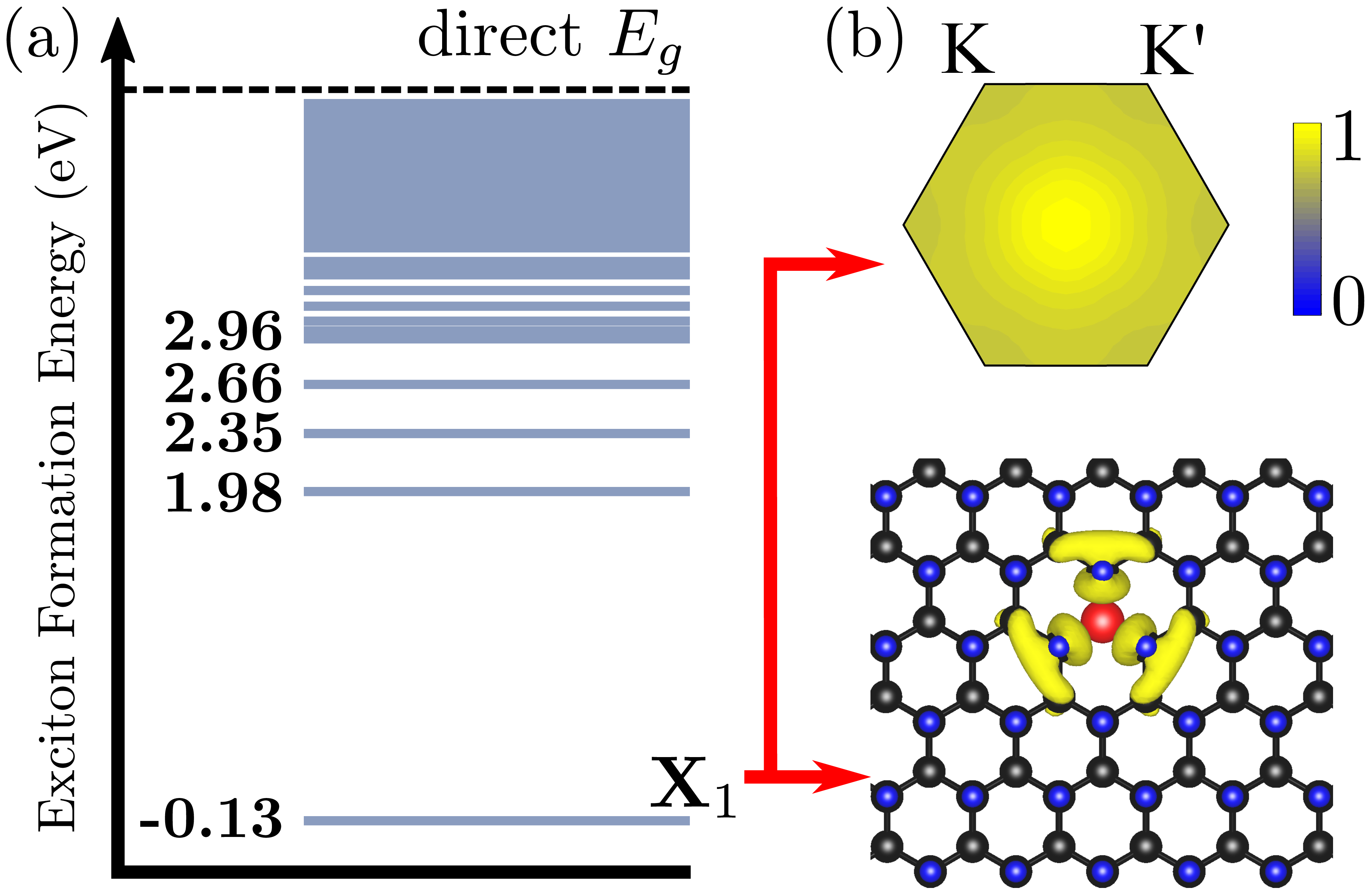}
\caption{\label{fig:fig2} (Color online) Low-energy excitation properties for \textbf{q}=0 excitons calculated by BSE@$GW_0$@PBE. (a) Energy spectrum below the direct band gap. Each horizontal line refers to an exciton state. A negative energy means spontaneous formation of the corresponding exciton. (b) Wavefunction plots of the $X_1$ exciton in the reciprocal (top) and real (bottom) space. The maximum value of wavefunction modulus has been renormalized to unity. In the real-space plot, the hole is fixed at the center (red ball).}
\end{figure}

Next, we calculate low-energy excitations of graphone by solving the BSE. First, let us consider the \textbf{q}=0 case. Table I lists the exciton formation energies ($E_f$'s) for the lowest-energy excitation ($X_1$) by different methods. They are all negative, which suggests that the graphone should form a many-body ground state by a spontaneous exciton condensation, namely, the formation of an excitonic insulator. Figure 2(a) shows the $E_f$'s of BSE@$GW_0$@PBE for all the direct transition excitons. Due to the spin-flipping, none of the excitons in this figure can be optically directly accessed. We should note that $E_f=-0.13$ eV for $X_1$ by $GW_0$@PBE is likely a bit too high, similar to $E_g$. The actual $E_f$ is expected to be more negative. Figure 2(b) shows for the $X_1$ exciton the reciprocal-space (top) and real-space (bottom) distributions. A large delocalization in the reciprocal space implies a high localization in the real space. Indeed, the $X_1$ exciton is localized to within $\sim$2 unit cells. Such a small exciton radius suggests that it is a boson, rather than a pair of (electron + hole) fermions.

In fact, here reported excitonic instability is insensitive to the hydrogenation coverage, provided that the two characteristic spin-polarized bands exist. For example, the calculation with one single H atom in a rectangular $1 \times \sqrt{3}$ and a $2 \times 2$ graphene supercell, which corresponds to a coverage of 25\% and 12.5\%, respectively, leads to $E_f=-0.44$ eV and $E_f=-0.59$ eV for the $X_1$ exciton by BSE@$GW_0$@PBE, consistently implying an even stronger tendency for excitonic instability.

A substrate is often required to support graphone for practical applications. To examine the role of the substrate, we place graphone on a monolayer hexagonal boron nitride (h-BN). Not only has h-BN been widely used for graphene, but also a strategy to grow graphone on h-BN has been suggested before\cite{Hemmatiyan}. Our calculations at the BSE@$G_0W_0$@PBE level show that the $E_f$ actually becomes more negative (-0.63 eV versus -0.55 eV) so the excitonic insulator phase becomes more stable. Despite being counterintuitive, one may understand this result as follows: an increase in screening due to the substrate leads to decreases in both the $E_g$ and the $E_b$, whose net difference determines the $E_f$. The smaller reduction of $E_b$, not in proportion to $E_g$, can be attributed to the fact that excitons in graphone are very small in size and hence are less affected by the low dielectric-constant h-BN substrate. Experimentally, the single-side hydrogenated graphene with coverage $>$20\% has been readily fabricated on the Ni\cite{Zhao,Bahn,Lizzit}, however, it needs to be transferred on a low dielectric-constant substrate like the h-BN to realize the spin-triplet excitonic insulator.

\begin{figure}[tbp]
\includegraphics[width=0.75\columnwidth]{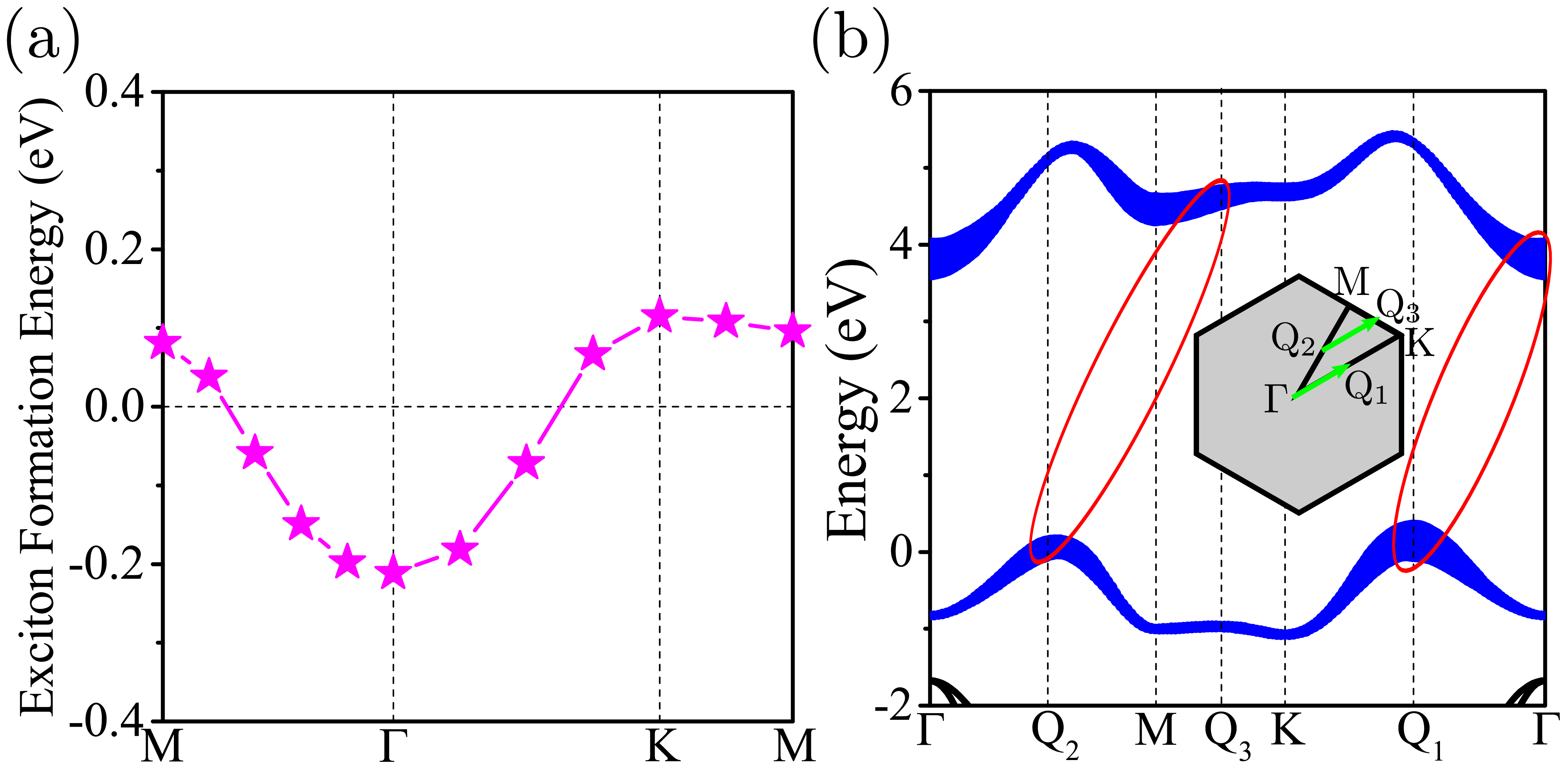}
\caption{\label{fig:fig3} (Color online) (a) \textbf{q}-dependent exciton formation energy along the high symmetric lines in the first Brillouin zone. A negative value means the spontaneous generation of excitons. The calculations were done using the Exciting code, which yields $E_f=-0.21$ eV at $\Gamma$, showing the same trend by Yambo with $E_f=-0.13$ eV. (b) The $k$-point specific contribution to the \textbf{q}= (1/6, 1/6) exciton wavefunction. The thicker the line, the larger the contribution. Shown in the inset are two vectors that meet the requirement on \textbf{q} along the boundary of the Brillouin zone, and the corresponding transitions are circled in red.}
\end{figure}

Second, we consider the \textbf{q$\neq$0} cases. For simplicity, Fig. 3(a) shows the exciton spectra for \textbf{q} along the high symmetric lines of the Brillouin zone. To our surprise, the \textbf{q}$\neq$0 excitons here are energetically less stable than the \textbf{q}=0 exciton, in spite of the fact that graphone has an indirect single-particle gap which is about 1-eV lower than the direct one. While a detailed analysis will follow, here we would like to point out that a reduced electric screening in 2D systems is likely the primary reason for the characteristics of the \textbf{q}-dependence for the excitons\cite{Cudazzo}. Moreover, the exciton dispersion over the Brillouin zone can be as large as 0.3 eV, which is not seen in three-dimensional materials. To understand these counterintuitive results, we write the exciton wavefunction as a linear combination of the electron-hole pair states
\begin{equation}\label{(1)}
\Psi_{\textbf{q}}(r_h, r_e) = \sum_{vc\textbf{k}}A^\textbf{q}_{vc\textbf{k}}\psi_{v,\textbf{k}}(r_h)\psi^{*}_{c,\textbf{k}+\textbf{q}}(r_e).
\end{equation}
where $\psi_{v,\textbf{k}}(r_h)$ and $\psi_{c,\textbf{k}}(r_e)$ are the wavefunctions of hole and electron, respectively. This allows us to define a quantity to measure the relative contribution of $\psi_{v,\textbf{k}}$ and $\psi_{c,\textbf{k}}$ at a given \textbf{k} to $\Psi_{\textbf{q}}$: namely, $\zeta^\textbf{q}_{v(c),\textbf{k}}=\sum_{c(v)}|A^\textbf{q}_{vc\textbf{k}}|$$^2$, with the sum over the conduction(valence) band index.

Figure 3(b) denotes the calculated $\zeta^\textbf{q}_{v(c),\textbf{k}}$ for \textbf{q}= (1/6, 1/6) as the width of the topmost single-particle valence band (v) and the bottommost single-particle conduction band  (c) at a given \textbf{k}. From inset of Fig. 3(b) (the Brillouin zone), it is clear that the transition between $\Gamma$ and $Q_1$ (defined as the midpoint between $\Gamma$ and $K$) and the transition between $Q_2$ (defined as the midpoint between $\Gamma$ and $M$) and $Q_3$ (defined as the midpoint between $M$ and $K$) both satisfy \textbf{q}= (1/6, 1/6). Indeed, at this \textbf{q}, in addition to the usual band extrema transition between $\Gamma Q_1$, the $Q_2 Q_3$ transition also contributes significantly as shown by red circles in Fig. 3(b), despite that the single-particle gap at $Q_2 Q_3$ of 4.6 eV is about 1 eV higher than that of $\Gamma Q_1$ of 3.59 eV\cite{notegap}. This indicates that one must consider the nonlocal effects from the {\it entire Brillouin zone} when calculating the exciton formation energy, as Fig. 3(b) is just one example of many possibilities which collectively lead to the indirect-to-direct crossover.

The indirect-to-direct crossover from single-particle band to many-body state due to the formation of excitons, nevertheless, may have important experimental consequences. For example, one may apply an electric field to alter the dielectric screening, whereby changing the exciton dispersion and subsequently measure the crossover in absorption. In addition, the screening is sensitive to an external magnetic field in magnetic systems, as well as being sensitive to an applied strain in general. Hence, applying any such an external stimulus or their combinations may induce a crossover to be measured. Experimentally, the exciton dispersion like the one in Fig. 3(a) can be measured by techniques such as the momentum-resolved electron energy loss spectroscopy (EELS) \cite{Egerton,Wachsmuth} or the resonant inelastic x-ray spectroscopy (RIXS) \cite{Schulke,Ament}, provided that the excitons have not condensed into a Bose condensate. On the other hand, if the excitons do form the condensate, they would only occupy the $\Gamma$ point in Fig. 3(a) and hence is dispersionless. This qualitative difference due to condensation offers an experimental means to distinguish the excitonic insulator from a band insulator in graphone.

Note that the exciton condensation here is fundamentally different from the one proposed in the literature for graphene\cite{Stroucken}. Graphene has a zero $E_g$ characterized by a Dirac-cone band structure. As such, the inclusion of an electron-hole interaction, regardless of its strength, will open up a many-body $E_g$ that drives the system into an excitonic insulator. The weak electron-hole coupling leads to a condensate of BCS-type\cite{Du,Lu}. In contrast, graphone exhibits a Mott-like insulating band structure with a more-than-3 eV spin-splitting $E_g$. The screening is very weak and as such there is a strong electron-hole coupling, giving rise to a condensate of BEC-type, instead\cite{Du,Lu}.

\begin{figure}[tbh]
\includegraphics[width=0.75\columnwidth]{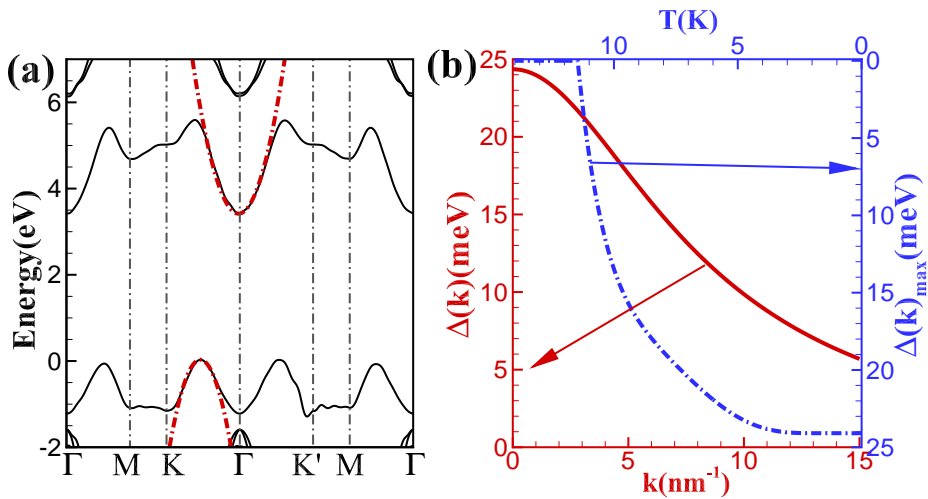}
\caption{\label{Fig4} (Color online) (a) The fitted \textbf{k}$\cdot$ \textbf{p} band structure of the graphone (red dashed lines). Black solid lines are the results from $GW_0$@PBE calculations. (b) The gap function $\Delta\left(\mathbf{k}\right)$ (see red solid lines) and temperature-dependence of the maximum gap $\Delta\left(\mathbf{k}\right)_{\max}$ (see blue dashed lines). See the text for more details.}
\end{figure}

To further understand the excitonic insulator phase, we model the graphone by an effective Hamiltonian in second quantization form as
\begin{align}
\hat{H}  &  =\sum_{\mathbf{k}}\left[  \epsilon_{a}\left(  \mathbf{k}\right)
-\mu\right]  a_{\mathbf{k}}^{\dag}a_{\mathbf{k}}+\sum_{\mathbf{k}}\left[
\epsilon_{b}\left(  \mathbf{k}\right)  -\mu\right]  b_{\mathbf{k}}^{\dag
}b_{\mathbf{k}}\nonumber\\
&  +\frac{1}{2}\sum_{\mathbf{k}^{\prime},\mathbf{k,q}}V_{aa}\left(
\mathbf{q}\right)  a_{\mathbf{k+q}}^{\dag}a_{\mathbf{k}^{\prime}\mathbf{-q}%
}^{\dag}a_{\mathbf{k}^{\prime}}a_{\mathbf{k}}\nonumber\\
&  +\frac{1}{2}\sum_{\mathbf{k}^{\prime},\mathbf{k,q}}V_{bb}\left(
\mathbf{q}\right)  b_{\mathbf{k+q}}^{\dag}b_{\mathbf{k}^{\prime}\mathbf{-q}%
}^{\dag}b_{\mathbf{k}^{\prime}}b_{\mathbf{k}}\\
&  -\sum_{\mathbf{k}^{\prime},\mathbf{k,q}}V_{ab}\left(  \mathbf{q}\right)
a_{\mathbf{k+q}}^{\dag}b_{\mathbf{k}^{\prime}\mathbf{-q}}^{\dag}%
b_{\mathbf{k}^{\prime}}a_{\mathbf{k}}\nonumber
\end{align}
where $\mu$ is the chemical potential and $\mathbf{k}$ is the in-plane momentum. $\epsilon_{a}\left(\mathbf{k}\right)$ and $\epsilon_{b}\left(\mathbf{k}\right)$ are the single-particle electron and hole energies, as depicted by the red dashed lines in Fig. \ref{Fig4}(a). $a_{\mathbf{k}}$ ($a_{\mathbf{k}}^{\dag}$) and $b_{\mathbf{k}}$ ($b_{\mathbf{k}}^{\dag}$) are the destruction (creation) operators of electron and hole. $V\left(\mathbf{q}\right)  $ is their mutual interaction, i.e., $V_{aa}\left(\mathbf{q}\right)=V_{bb}\left(\mathbf{q}\right)=V_{ab}\left(\mathbf{q}\right) =\frac{e^{2}}{2\varepsilon\left\vert \mathbf{q}\right\vert }$, where $\varepsilon$ is a screening parameter. We take it here as the in-plane dielectric constant of the graphone, namely, $\varepsilon=4.6\varepsilon_0$ by deducing from the first-principles calculations\cite{note46}. To solve this many-body interaction problem, following ref. \cite{Zittartz1967}, we first use the mean field approximation and diagonalize the pair Hamiltonian by introducing a Bogoliubov transformation of $H$, which yields a numerical (not operator) term in the Hamiltonian. Then, minimizing this term with respect to the transformation constant leads the generalized coupled equations%
\begin{align}
\Delta\left(  \mathbf{k}\right)   &  =\sum_{\mathbf{k}^{\prime}}%
V_{\mathbf{k-k}^{\prime}}^{ab}\frac{\Delta\left(  \mathbf{k}^{\prime}\right)
}{E\left(  \mathbf{k}\right)  }\left[
\begin{array}
[c]{c}%
1-f\left(  \frac{E\left(  \mathbf{k}^{\prime}\right)  -\eta\left(
\mathbf{k}^{\prime}\right)  }{2}\right) \\
-f\left(  \frac{E\left(  \mathbf{k}^{\prime}\right)  +\eta\left(
\mathbf{k}^{\prime}\right)  }{2}\right)
\end{array}
\right] \\
\xi\left(  \mathbf{k}\right)   &  =\epsilon_{{\normalsize P}}\left(
\mathbf{k}\right)  -\mu\nonumber\\
&  -\sum_{\mathbf{k}^{\prime}}V_{\mathbf{k-k}^{\prime}}^{ab}\left[
1-\frac{\xi\left(  \mathbf{k}^{\prime}\right)  }{E\left(  \mathbf{k}^{\prime
}\right)  }\right]  \left[
\begin{array}
[c]{c}%
1-f\left(  \frac{E\left(  \mathbf{k}^{\prime}\right)  -\eta\left(
\mathbf{k}^{\prime}\right)  }{2}\right) \\
-f\left(  \frac{E\left(  \mathbf{k}^{\prime}\right)  +\eta\left(
\mathbf{k}^{\prime}\right)  }{2}\right)
\end{array}
\right] \\
\eta\left(  \mathbf{k}\right)   &  =\epsilon_{{\normalsize M}}\left(
\mathbf{k}\right)  -\mu\nonumber\\
&  -\sum_{\mathbf{k}^{\prime}}V_{\mathbf{k-k}^{\prime}}^{ab}\left[
\begin{array}
[c]{c}%
1-f\left(  \frac{E\left(  \mathbf{k}^{\prime}\right)  -\eta\left(
\mathbf{k}^{\prime}\right)  }{2}\right) \\
+f\left(  \frac{E\left(  \mathbf{k}^{\prime}\right)  +\eta\left(
\mathbf{k}^{\prime}\right)  }{2}\right)
\end{array}
\right],
\end{align}
where $\epsilon_{{\normalsize P}}\left(\mathbf{k}\right)$ and $\epsilon_{M}\left(\mathbf{k}\right)$ are defined as $\epsilon_{P}\left(\mathbf{k}\right)=\epsilon_{b}\left(\mathbf{k}\right)+\epsilon_{a}\left(\mathbf{k}\right)$ and $\epsilon
_{M}\left(\mathbf{k}\right)=\epsilon_{b}\left(\mathbf{k}\right)-\epsilon_{a}\left(\mathbf{k}\right)$, respectively, and $f\left(x\right)=1/\left(e^{x}+1\right)$ is the Fermi distribution function. $E\left(\mathbf{k}\right)$ is the pair-breaking excitation spectrum: the energy cost to replace a bound exciton in the condensate with an unbound electron-hole pair in plane-wave states of momentum $\mathbf{k}$. From the relation $\Delta^{2}\left(\mathbf{k}\right)=E^{2}\left(\mathbf{k}\right)-\xi^{2}\left(\mathbf{k}\right)$, we can obtain the gap function $\Delta\left(\mathbf{k}\right)$ by solving the above coupled equations self consistently. In practice, for simplicity, we deal with the indirect transition from the valence band maximum to the conduction band minimum. It corresponds to the formation of excitons with a momentum transfer \textbf{q}= (1/6, 1/6). However, this does not change the basic physical picture, as this indirect transition also has a negative $E_f=-70$ meV [see Fig. 3(a)].

As depicted by the red solid line in Fig. \ref{Fig4}(b), a many-body gap $\Delta\left(\mathbf{k}\right)_{\max}=$24.1 meV is found for the graphone, which is on the same order of magnitude with the energy gain (70 meV) by first-principles $GW$-BSE calculations. We also calculate the temperature dependence of $\Delta\left(\mathbf{k}\right)_{\max}$ and the result is shown by the blue dashed line in Fig. \ref{Fig4}(b). A critical temperature of $T_{c}=$11.5 K is found, above which the many-body gap is closed and hence the excitonic insulator phase is destroyed. Note that for such a many-body system, while $T_c$ correlates with $E_f$, the relation is not as simple as $T_c \sim E_f/k_B$.

In summary, first-principles calculations, combined with effective Hamiltonian modeling, predict graphone as a spin-triplet excitonic insulator. The great advantage for uncovering this class of materials is their easiness to be unambiguously identified experimentally, as their spin characteristics make it possible to circumvent the current difficulties of using conventional probes for Bose condensation. In particular, the super flow of the exciton condensate can now be measured by means of a transport experiment. Moreover, the presence of a crossover between direct-gap exciton spectrum and indirect-gap single-particle band, as well as the unusually large exciton dispersion, is advantageous in terms of modulating the low-energy excitation. It is also possible to tune the low-energy exciton characteristics by altering the system's dielectric screening through an applied external field.

\begin{acknowledgments}
Work in China was supported by the Basic Science Center Project of NSFC (Grant No. 51788104), the Ministry of Science and Technology of China (Grant Nos. 2016YFA0301001 and 2018YFA0306101), the National Natural Science Foundation of China (Grant Nos. 11674071, 11674188, 11434010, 11974340, 11574303 and  61674145), the Beijing Advanced Innovation Center for Future Chip (ICFC), the Open Research Fund Program of the State Key Laboratory of Low-Dimensional Quantum Physics (NO. KF201702), and the Beijing Institute of Technology Research Fund Program for Young Scholars. Work in the US (SBZ) was supported by US DOE under Grant No. DE-SC0002623. SBZ had been actively engaged in the design and development of the theory, participated in all the discussions and draft of the manuscript.
\end{acknowledgments}


\begin{thebibliography}{90}
\bibitem{notegap} A scissor operator is used to correct the direct gap to the $GW_0$ values of the VASP which makes the indirect gap slightly different.
\bibitem{note46} We define an effective in-plane 2D dielectric constant as $\varepsilon = 1 + 4\pi\alpha_{\rm 2D}/d$, where $\alpha_{\rm 2D}$ and $d$ are the 2D polarizability and the thickness measure of the out-of-plane charge distribution of graphone, respectively. See Ref. \onlinecite{usPRL} for more details about the $\alpha_{\rm 2D}$.
\end{thebibliography}
\end{document}